\documentclass[pre,aps,floatfix,superscriptaddress,onecolumn]{revtex4-2}
\usepackage{graphicx}
\usepackage{dcolumn}

\usepackage{latexsym}
\usepackage{hyperref}
\usepackage{amsmath, amsthm, amssymb}
\usepackage{epsfig}
\usepackage{bm}
\usepackage[dvipsnames]{xcolor}

\begin{document}

\title{Synchronous and Asynchronous Updates of Active Ising Spins in One Dimension}
\author{Anish Kumar}
\email[]{anishkumar.rs.phy22@itbhu.ac.in}
\affiliation{Department of Physics, Indian Institute of Technology(BHU), 
Varanasi- 221005, India}

\author{Sudipta Pattanayak}
\email[]{sudipta.pattanayak@cyu.fr}
\affiliation{Collège de France, Université Paris Sciences et Lettres,  Paris, France}.
\affiliation{Institut Curie, Université Paris Sciences et Lettres, Physique de la Cellule et cancer, UMR 168, 
Paris, France
}

\author{R. K. Singh}
\email[]{rksinghmp@gmail.com}
\affiliation{Department of Physics, Bar-Ilan University, Ramat-Gan 5290002, Israel}

\author{Shradha Mishra}
\email[]{smishra.phy@itbhu.ac.in}
\affiliation{Department of Physics, Indian Institute of Technology(BHU), 
Varanasi- 221005, India}

\begin{abstract}

    How do update rules affect the dynamical and steady state properties of a flock? In this study, we have explored the active Ising spins $(s = \pm 1)$  in one dimension, where spin updates its orientation according to the Metropolis algorithm (based on the neighbors) via two different update rules. (i) Parallel, and (ii) Random-sequential. We explore the effect of Parallel and Random-sequential updates on the dynamical properties 
    of flocks in one dimension. Due to the inherent asynchronous nature of the Random-sequential update, the directional switching of the flock is increased compared to the Parallel one. The nature of phase transition is affected by the difference in the updating mechanism: discontinuous 
    for Parallel and continuous for Random-sequential updates. %{\bf discuss the mechanics behind the first order and continuous transition} 
\end{abstract}

\maketitle

\section{Introduction}
Collective motion is a ubiquitous phenomenon observed in active systems 
driven out of equilibrium across widely separated length scales from 
single cells \cite{kemkemer2000nematic}, to unicellular organisms \cite{bonner1998way}, to 
bird flocks \cite{parrish1997three}, and human crowd \cite{helbing2000simulating}. The emergence of such a motion in a 
group of self-propelled units is termed as a flocking transition \cite{vicsek2001question,pattanayak2018collection} and 
was reported by Vicsek and coworkers for a system of particles in two dimensions 
\cite{vicsek1995novel}. While most natural systems of interest are two or three-dimensional, the formation of collective motion in one dimension has gained attention in recent years \cite{czirok1999collective,solon2013revisiting}. Such one dimensional 
flocks exhibit the interesting property of direction switching \cite{o1999alternating,dossetti2011cohesive} 
and recent theoretical and experimental studies have proven the usefulness of the study of 
collective motion in one dimension, in particular relevance to the phenomena of directional 
switching \cite{yates2009inherent,buhl2006disorder}. 

Most such models studying flocking in one dimension generally employ discrete time evolutions 
of the system, which are closer in nature to the sense of time as represented in digital simulations. 
However, it is known that on a digital time scale, a system of multiple particles can exhibit 
properties that are not a true representation of the original dynamical system, as has been 
observed in equilibrium \cite{choi1983digital} and nonequilibrium systems \cite{blok1999synchronous}. Such differences 
in update rules have led to the appearance of new universality classes in coupled map 
lattices \cite{marcq1997universality,rolf1998directed}. The observations motivate us to study the effects of different 
update rules on flocking dynamics in a collection of self-propelled particles. In order to proceed 
with our goal, we introduce a system of active Ising spins moving in the unit interval $[0, 1]$ 
with constant speed $v_0$, which interact locally in a neighborhood of range $\delta x$. We find 
that the differences in update rules reflect in both the transient and steady state properties 
of the flocks. The nature of phase transition also changes with the change in the update rule. The paper is organized as follows: 
The details of the two update rules are covered in section \ref{II}. Following that, section \ref{III}  discusses how the update rules change the system's properties, and section \ref{IV} presents the conclusion.

%%%%%%%%%%%%%%%%%%%%%%%%%%%%%%%%%%%%%%%%%%%%%%%%%%%%%%%%%%%%%%%%%%%%%%%%%%%%%%%%%
\section{Comparison of the two update rules}\label{II}
In this work, we consider N active Ising spins ($s = \pm 1$) moving along a unit interval [0,1] with a fixed 
speed $v_0$ with periodic boundary conditions. Initially, all the spins are randomly distributed within the unit interval. Each $i^{th}$ spin $s_i$ interacts with all other spins within the interval $[x_i-\delta x, x_i+\delta x]$ and updates its spin orientation according to the Metropolis algorithm \cite{newman1999monte}. Where $\delta x$ is the interaction range and $x_i$ is the position of $i^{th}$ spin. If $f$ is the net spin
within the interaction range $\delta x$, then the spin updates its orientation depending on the product $s_i f$.
If the product is negative, the spin is certainly flipped, and if the product is positive, then the spin is flipped with probability  $\exp(-\beta s_if)$, where  $\beta$ is the inverse temperature.
Each $i^{th}$ spin updates its position according to the given equation:
\begin{align}
\label{dyn}
\tilde x_i = x_i + v_0 \tilde s_i
\end{align}
where $\tilde x_i$, $\tilde s_i$ is the updated position and orientation of $i^{th}$ spin respectively.
To observe the effect of the update rule on the dynamical properties of the system, we have updated the above system with two different update rules. (i) Parallel and (ii) Random-sequential update.\\
In the Parallel update rule, each spin updates its orientation ($s_i \rightarrow  \tilde s_i$) based on the Metropolis algorithm. Once the orientation of all the spins is updated by $\tilde s_i$, then everyone's position is updated simultaneously using Eq. (\ref{dyn}). This counts as the one Monte-Carlo step for Parallel update.
In contrast, an active spin $i$ is randomly selected from the set of $N$ spins in the Random-sequential update, and its spin $s_i$ is modified using the Metropolis method, and then its position is updated in accordance with (\ref{dyn}). The updated value of spin $s_i$ is applied right away to change the position of the $i^{th}$ particle. In order to give each spin an equal chance of updating, this process is done $N$ times. This $N$ random flip process makes up one unit of time that is equivalent to one Parallel update of $N$ spins. It may be concluded from the definitions of these update rules that the Random-sequential is asynchronous, whereas the Parallel update is synchronous. The Random-sequential updating rule has an underlying randomness. It also demonstrates a significant impact on the dynamics and steady state characteristics of the system.\\

To characterise the orientational ordering among the spins we defined $m$ (Net magnetization) as the order parameter of the system. $m$ is defined as:
\begin{align}
\label{order}
m = \frac{1}{N}\sum_i s_i\, 
\end{align}
The magnitude $|m|$ takes values in the interval $(0, 1)$ with 0 representing a completely disordered state and 1 the state of a completely flocking state. The other parameters in the system: the self-propulsion speed $v_0$ is varied from $[0.001, 0.003]$ and interaction range $\delta x$ is varied from $[0.01, 0.05]$ and inverse temperature $\beta $ from $[1,5]$. The system is studied for $N=500, 1000$ number of spins. The total simulation time step is $5 \times 10^5$. The thermal averaging is performed over approximately $100$ independent realisations for better statistics. We started with an initially random arrangement of spins with random orientation. With time, the system evolves and reaches a steady state determined by the parameters. The steady state is defined when the characteristics of the system remain statistically the same with respect to time.
%On the other hand, in the Random-sequential update, an active spin $i$ is chosen randomly from the collection 
%of $N$ spins, and its spin $s_i$ is modified according to the Metropolis algorithm. The difference lies in the 
%step that the updated value of spin $s_i$ is 
%used immediately to modify the position of the $i^{th}$ particle according to 
%(\ref{dyn}). This process is repeated $N$ times so that each spin gets an equal chance 
%of update, and this process of $N$ random flips constitutes one unit of time equivalent 
%to a Parallel update of $N$ spins. 
%{\color{red}{Here, at a single time step, once a spin state is updated, its modified state will be used for other spins if it lies within its neighborhood.}} 

%It is evident from the definition of the two update rules 
%that a Parallel update is synchronous, whereas a Random-sequential update is intrinsically 
%asynchronous, as there is a randomness inherent in the very nature of the update rule. 
%Such randomness has implications in the dynamics of the spin system and reflects 
%in both the transient and steady state properties. 

%At $t = 0$, the positions of the spins are chosen uniformly from the unit interval. Initial spin states are also 
%chosen $\pm 1$  at random in all the following observations unless explicitly stated. 
\section{Results}\label{III}
\begin{figure}
\includegraphics[width=0.7\textwidth]{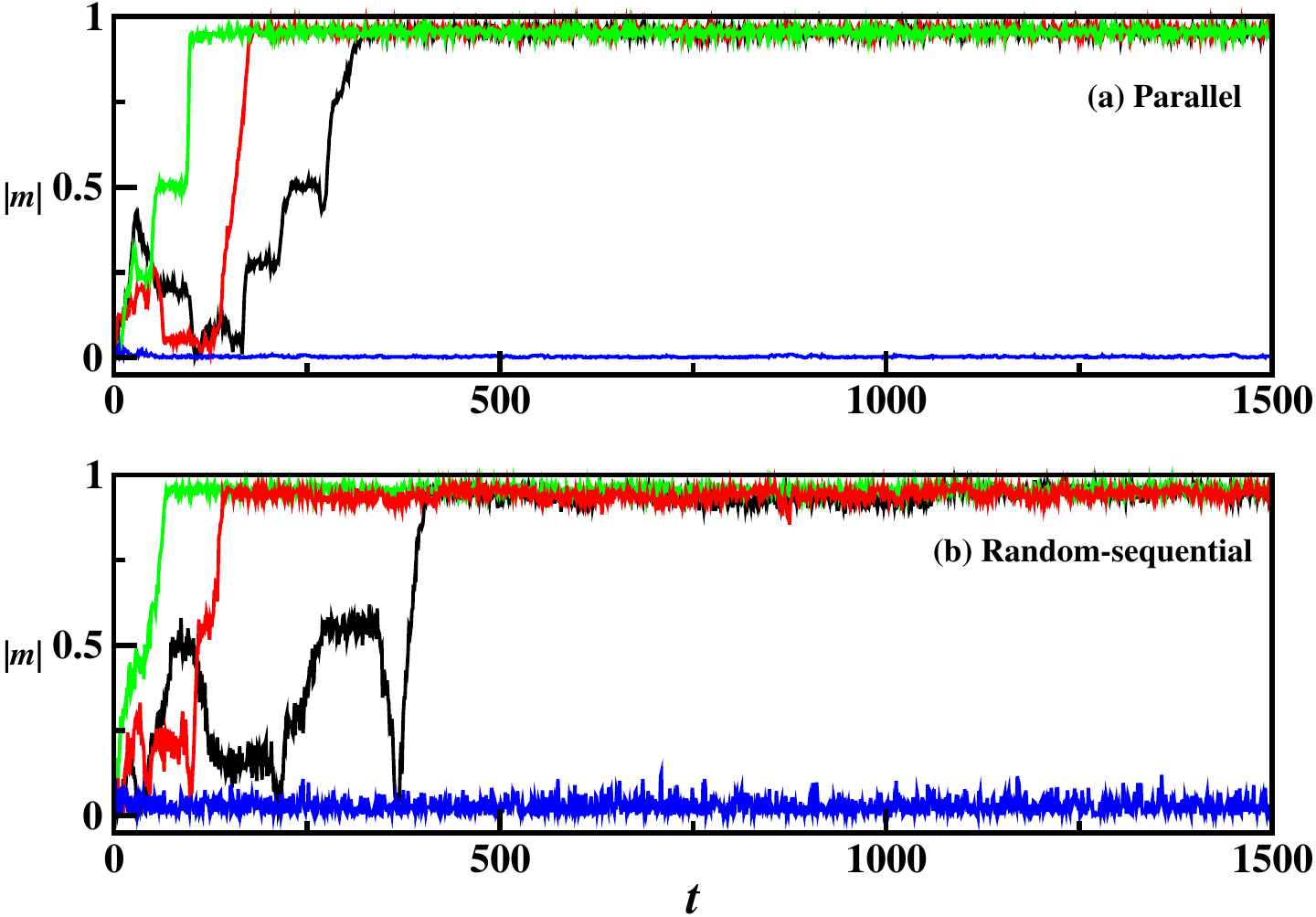}
\caption{(color online) Variation of the order-parameter $ |m|$ with time $t$  for a system 
	of $N = 1000$ spins for Parallel (a) and Random-sequential (b) updates. Parameter values for the 
	observations are: $(v_0, \delta x) =  (0.001, 0.01) $
    (black), $(0.001, 0.05)$ (red) and $(0.003, 0.01)$
     (green)  
	respectively for inverse temperature $\beta = 4$. Blue is for $\beta = 1$ and $(v_0, \delta x) = (0.001, 0.01)$.  }
\label{fig1}
\end{figure}

In Fig.~\ref{fig1}, we have shown the variation of the order-parameter $|m|$ time series for a 
system of $N = 1000$ spins moving in the unit interval $[0,1]$ for different $(v_o,\delta x)$ combinations for Parallel (a) and Random-sequential (b) updates respectively. For small inverse temperature $\beta = 1$ starting from the random initial state $|m| \simeq 0$, the system remains in the disordered state with $|m| \simeq 0$, whereas for large $\beta = 4$, it reaches to the ordered state with $|m| \simeq 1$ in the steady state. \\
%For large inverse temperatures, e.g., $\beta = 4$, 
The $|m|$ vs. $t$ curves are similar for the two update rules in the steady state for $\beta=4$. As the $v_0$ or $\delta x$ is increased, the system takes less time to reach the ordered state.
However, for $\beta = 1$ (blue), we observe relatively larger fluctuations in the order parameter for the Random-sequential update compared to the Parallel one. These are the consequences of the inherent randomness in the Random-sequential update rule. \\ 
Starting with a random distribution of the spins at $t = 0$, the spins tend to move with the spins of the same $s$ when the system exhibits 
long-range order, implying that the emergence of long-range order for high $\beta$ values is a mean-field effect. The cause for this effect is the propagation of the local interaction amongst the spins 
across the interval due to the finite speed of movement of the spins, i.e., $v_0 > 0$. Such propagation of the local interaction leads to the emergence of a long-range order when global fluctuations are 
less (high $\beta$ values). This is because for high $\beta$, the probability of flipping  against the majority 
$\exp(-\beta s_i f)$ is less at any instant $t$, and hence the alignment of all the spins along the interval 
is achieved. On the other hand, for low values of $\beta$, the enhanced magnitude of global fluctuations increases 
the chance of any given spin $s_i$ to flip against the majority. As a result, even when the spins are moving 
with a constant speed $v_0$, long-range order is not established because of the increased strength of 
global fluctuations, which tends to disrupt the established local order at every instant. Later, we will show that the fluctuations are relevant not only for low values of $\beta$ but also for higher values.

\begin{figure}[h]
\includegraphics[width=0.7\textwidth]{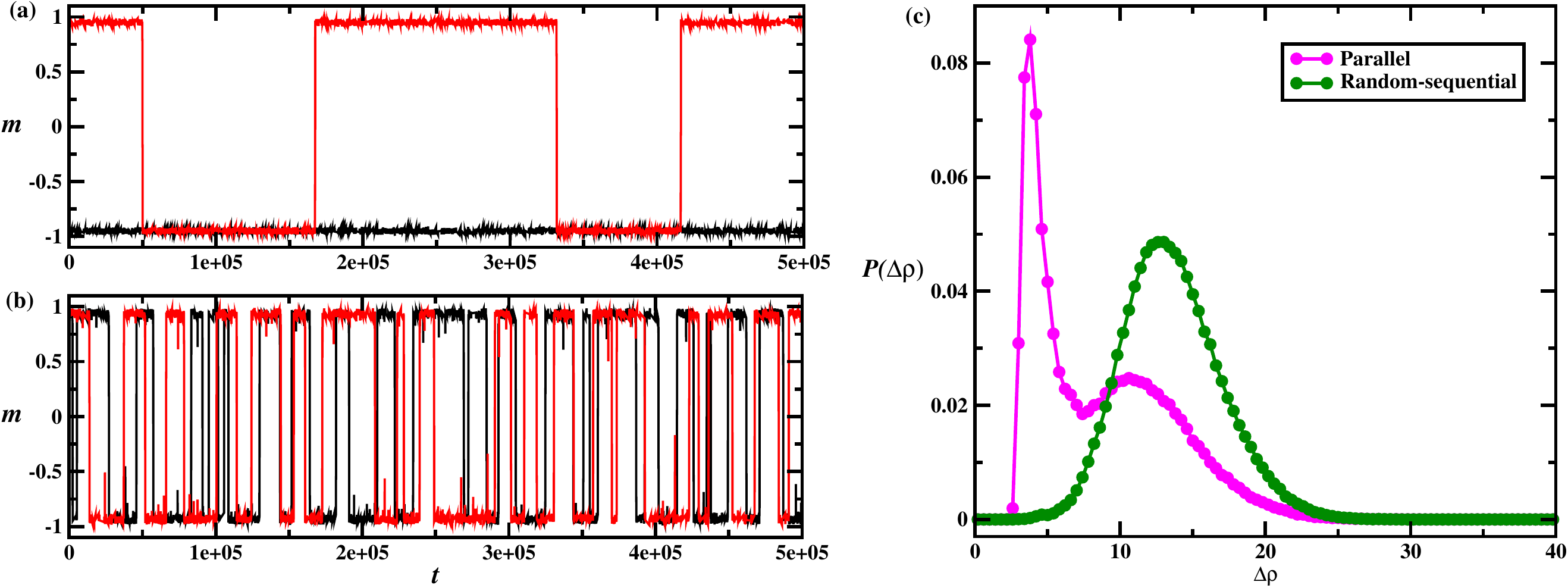}
\caption{(colour online) Plots (a-b) shows order-parameter $m$  time series for two different realizations, the Parallel and Random-sequential update, respectively. In Plot (c), probability distribution function (PDF) of density fluctuations  $P(\Delta {\rho}) $ for Parallel and Random-sequential updates. 
 The systems consist of 
	$N = 1000$ spins moving with speed $v_0 = 0.001$ at inverse temperature $\beta = 4$. 
	The interaction range of the spins $\delta x = 0.01$. Different colours in the plots (a-b) represent different realizations. Distribution is calculated using 500000 iterations over 100 different realizations.
}
\label{fig2}
\end{figure}
%{\color{red}{To further demonstrate the system properties in the ordered region Fig.~\ref{fig2}(a-b) shows the snapshots of the number of spins (local density $P(x)$) between positions $(x - \delta x, x + \delta x)$ of the spins 
%for the two update rules for $\beta = 4$ for Parallel and Random-sequential updates respectively.  
%The Random-sequential updates have additional randomness due to the asynchronous updating mechanism. Due to that a given $i^{th}$ spin updates its orientation with respect to its neighbours instantaneous updated orientation, whereas in the Parallel update the same spin still seeing the previous step orientation of its neighbours, while updating its own. This difference in mechanism of update lead to prompt response with respect to change of neighbours orientation. Which can further results into spins to be in the close proximity of spins of same orientation for high $\beta$, hence denser clusters as can be seen from Fig. \ref{fig2}(b) at later times. In comparison for the Parallel update there is a delayed response to the neighbours lead to the more homogeneous clusters for large $\beta$ values at late times as well as shown in Fig.~\ref{fig2} (a). }} 

To further demonstrate the system properties in the ordered region, Fig.~\ref{fig2}(a-b) shows order parameter $m$ time series for two different realizations for the Parallel and Random-sequential update for $(v_0, \delta x) = (0.001, 0.01)$ and high $\beta = 4.0$. For both updates, the system shows the globally ordered state with $|m| \simeq 1$. But the directions of spins globally switches from one orientation type to another, hence $m$ shows switching from $+1$ to $-1$ state.  The frequency of switching is higher for the Random sequential update in contrast to the Parallel update. The two independent realisations shown by black and red are equivalent for Random sequential update, whereas the black shows zero switching and red shows the few numbers of switching for Parallel update. The synchronous nature of dynamics for the Parallel update makes the global switching more costly in comparison to Random-sequential, where each spin can flip randomly and slowly can turn the whole system to reorient its direction. Further, this leads to a stronger global orientation for Parallel updates. Now, we look at how the distribution of the density of spins in the system leads to the difference in the fluctuations in the magnetisation. We analyse the distribution of spins in space for the two updates and find that for the Parallel update, spins are either homogeneously arranged in the space with the same orientation or a strong density band with a globally ordered state. For the Random sequential update, the ordered state always results in the clustering of spins in the system. This motivated us to calculate the density fluctuations in the system for the two updates.  In Fig.~\ref{fig2}(c), the probability distribution function (PDF) of density fluctuations $P(\Delta \rho)$ for Parallel and the Random-sequential updates is shown. The other parameters are the same as in (a-b). The density fluctuation is defined as  $\Delta {\rho}(t)  = \sqrt{\langle \rho(t)^2 \rangle -\langle \rho(t) \rangle^2}$. The local density $\rho(t)$ is obtained by dividing the whole system in smaller bins (coarse-grained regions) and calculating the number of particles in the coarse-grained region. Further $\langle ... \rangle$ mean average over all the bins. It is calculated as a function of time and for different independent realisations. 
In the Random-sequential update, the system reverses its direction of motion frequently, so there is a dense cluster in the system that reverses its direction of motion. However, in the parallel update, due to the synchronous nature of the update, the reversal time is quite high. Sometimes, within the given time interval, it does not show any reversal, i.e. homogeneous distribution of the spins in the system. But occasionally it also switches the direction and dense clusters can also form. So, in the Parallel update case, we have found that the probability distribution function $P(\Delta \rho)$ shows a bimodality. But in the Random-sequential case, there is a single peak in the distribution.

Till now, we focused on the nature of the cluster of spins for the two updates. Further, we calculated the number fluctuation in the system to investigate the effect of the inherent randomness in the Random sequential case on the local density of particles. For comparison we calculate the number fluctuation for both the updates. To calculate the number fluctuation, we start with a small length at the center of the system and gradually increase the length and calculate the number fluctuation in the different lengths for both update rules. % Fig.~\ref{fig4}, shows the number fluctuation $\Delta N$ vs. $\langle N \rangle$ where $\langle N \rangle$ is the mean number of particles in the different lengths from the center of the system. 
We find that for both the cases $\Delta N \sim \langle N \rangle ^\alpha$. The  $\alpha = 1/2$ for the equilibrium systems \cite{beale2021statistical} and $\alpha \simeq 1$ for self-propelled particles \cite{ mishra2006active, chate2008collective, dey2012spatial,ramaswamy2003active,mishra2014aspects,singh2021ordering,jena2023ordering,semwal2024macro}. For the Parallel update, similar to the density fluctuation which is bimodal in nature as shown in Fig. \ref{fig2}(c),  the $\Delta N$ also shows the dual character. For some of the realisations, the density remains homogeneous and $\Delta N \simeq N^{\alpha}$ with smaller $\alpha$ and for many other realisations $\alpha \simeq 1$. Whereas for the Random-sequential update, the density fluctuation remains the same for all realisations. The difference over the realisations for Parallel update can also be obtained over time, but the time needed for the same is much larger than the maximum simulation time in our present study.  

\begin{figure}[h]
	\includegraphics[width=0.7\textwidth]{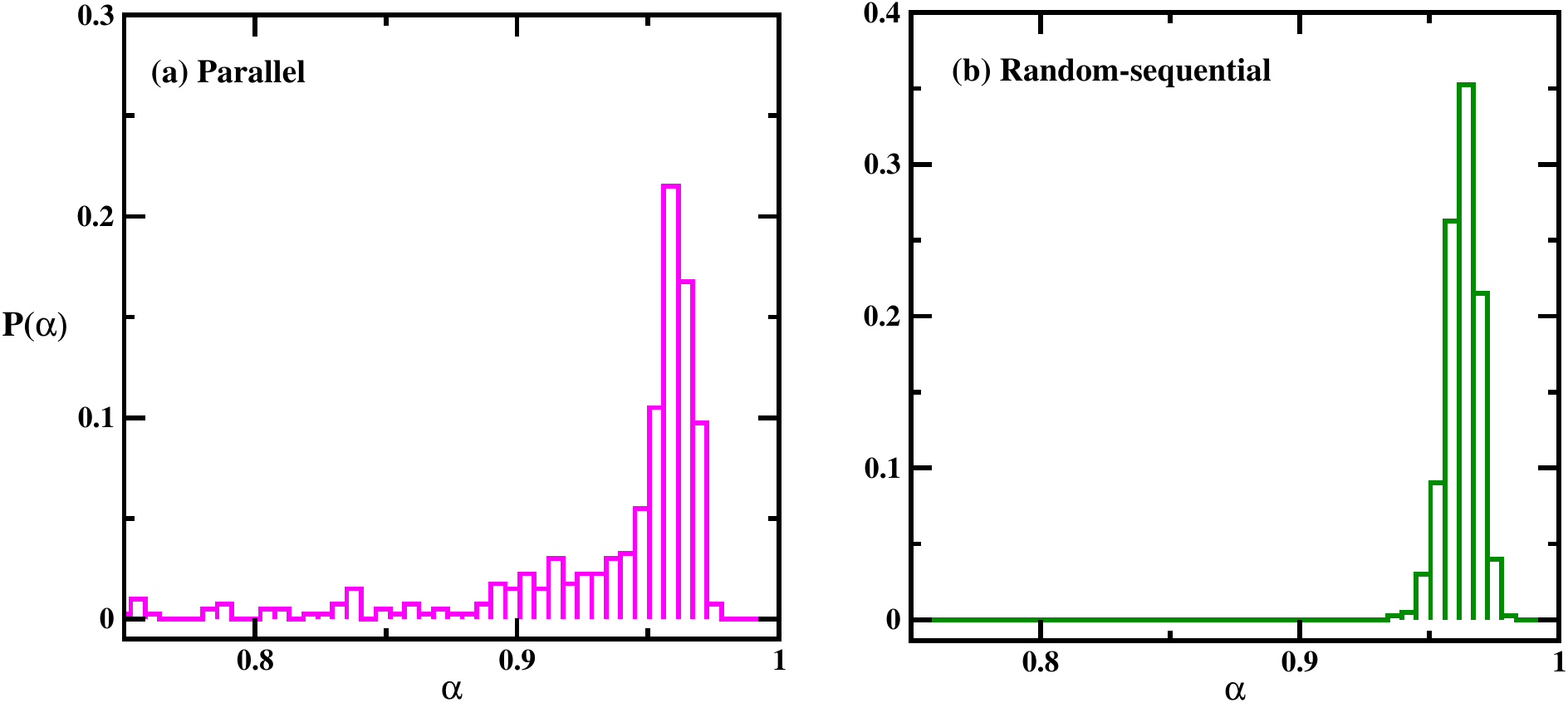}
	\caption{(color online) Plots (a) and (b) show probability distribution $P(\alpha)$ of the Number fluctuation exponent $\alpha$ in the ordered state for $N = 1000$ spins for Parallel and Random-sequential update, respectively. All other parameters are similar to Fig.~\ref{fig2}.  
	}
	\label{fig4}
\end{figure}

In Fig. \ref{fig4}(a-b), we show the probability distribution function (PDF) of the exponent $P(\alpha)$ is shown for Parallel and Random-sequential update, respectively. For the  Parallel update, 
the distribution shows the large variation of  $\alpha$ at the front of the distribution, whereas for the   Random-sequential update, it is unimodal with a single peak close to $\alpha \simeq 0.97$. Hence, the system shows the Giant number fluctuation (GNF) for both the updates.
It should be noticed that the GNF, reported in previous studies, is for a system of dimensions two and higher. The appearance of large density fluctuation in a one-dimensional system is truly remarkable. 

\begin{figure}[h]
\includegraphics[width=0.7\textwidth]{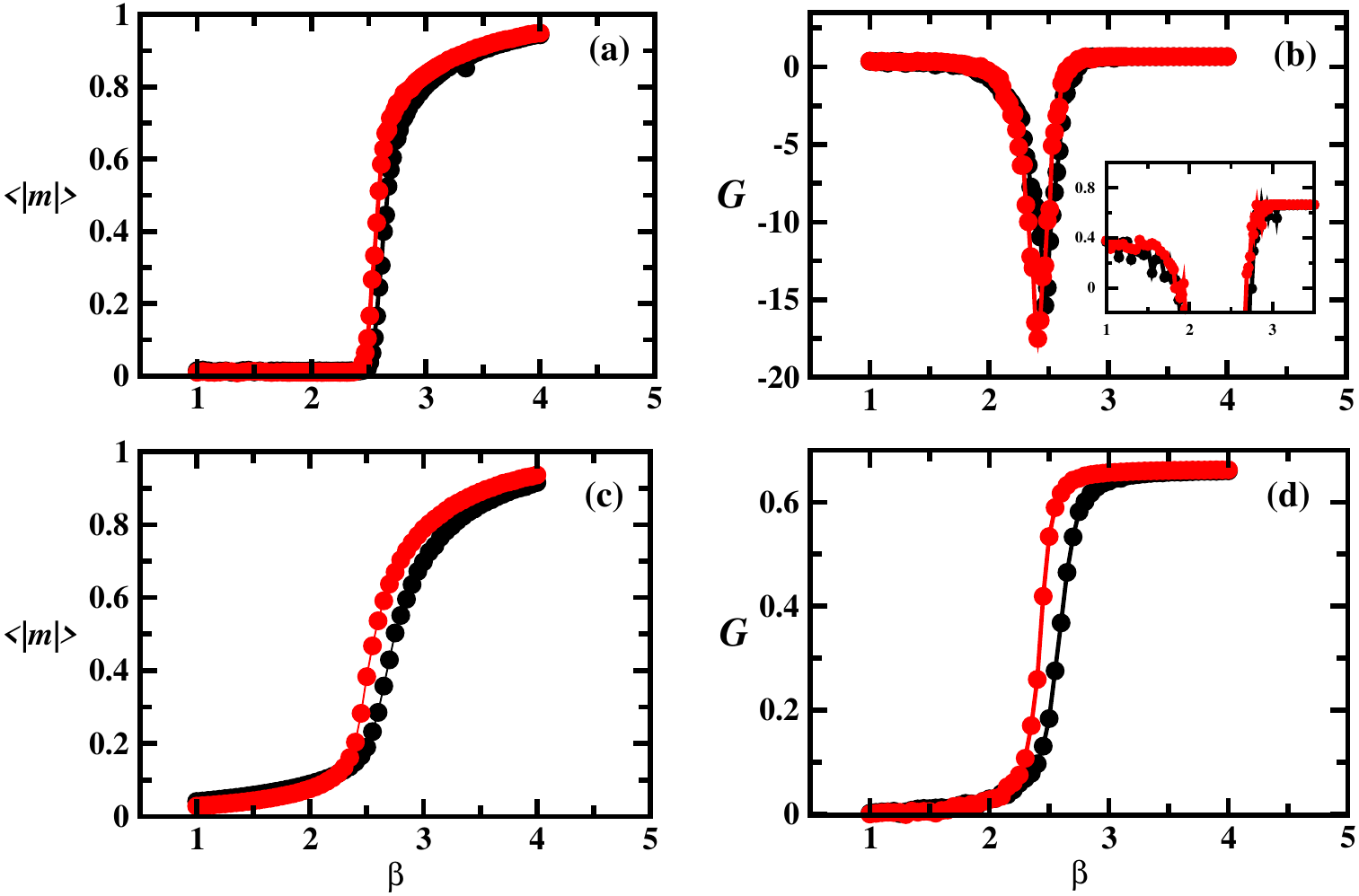}
\caption{(color online) Plots (a-b) show the variation of the order parameter $\langle|m|\rangle$ and the Binder cumulant $G$ in the steady state for 
Parallel and (c-d) for Random-sequential updates, respectively. The steady state properties are calculated using 
50000 iterations with 10000 iterations for the thermal averaging  and averaged over 100 ensembles. In (a-d), solid black circles are for $N = 500 $, and solid red circles are for $N = 1000$ spins. Other parameters are the same as in Fig.~\ref{fig2}.}
\label{fig5}
\end{figure}

\begin{figure}[h]
	\includegraphics[width=0.7\textwidth]{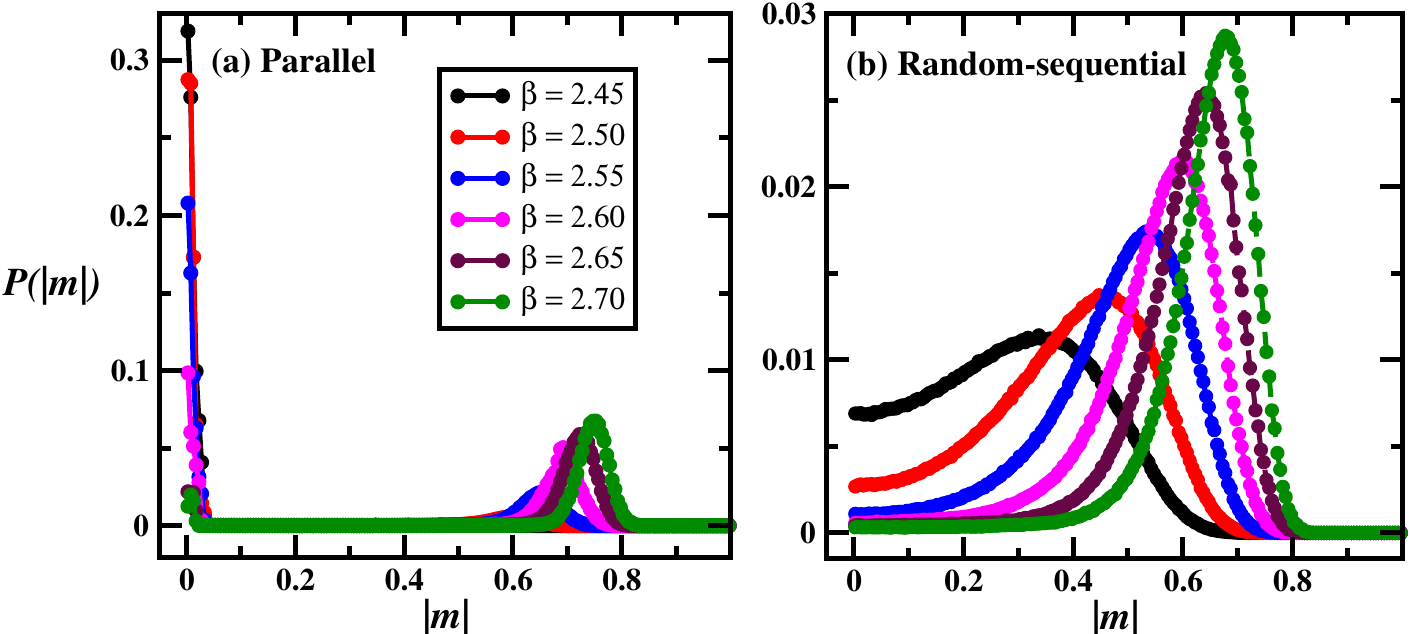}
	\caption{(color online) Plots (a) and (b) show the probability distribution $P(|m|)$ of the order parameter $|m|$ in the neighbourhood of transition 
$\beta$ for six different values $\beta$ for the two update rules.  
The bimodal nature of the order-parameter distribution (a) and unimodal character with 
means shifting towards the right with increasing $\beta$ (b).  The distributions are calculated using 50000 iterations 
over 100 ensembles. All other parameters are the same as in Fig.~\ref{fig2}. 
	}
	\label{fig6}
\end{figure}

The previous results here have focused on the steady state characteristics of the system for the two updates when the system is in the deep ordered state. 
Now, we discuss the effect of update rules when the system is close to the order-disorder transition point. For that, we vary the $\beta$ and explore the system near critical $\beta$.  The system shows the order-to-disorder phase transition as $\beta$ is varied from large to small values. We characterise the nature of phase transition for the two types of updates.
To check the effect of the update rules on the nature of phase transition, in Fig.~\ref{fig5}(a-c), (b-d), we show the order-parameter, fourth-order Cumulant (Binder Cumulant) for a range of inverse temperature $\beta$. The Binder Cumulant is defined by;
$G = 1-\langle |m|^4 \rangle/3 \langle |m|^2 \rangle^2$
is useful for characterizing the nature of phase transition in many nonequilibrium systems \cite{chate2008collective,solon2013revisiting,durve2016first,bhattacherjee2015topological,singh2021bond}. In the case of first-order transition, there is a sharp drop toward the negative value near the transition region.  For the case of Parallel update as shown in Fig.~\ref{fig5}(b), $G$ goes from the value $1/3$ for low $\beta$ value (disordered region) to $2/3$ value for large $\beta$ values (ordered region) Fig.\ref{fig2}(b)) with a sharp dip to negative values near transition. Accordingly, the order parameter $\langle |m| \rangle$ vs. $\beta$ plot shows the jump around the order-disorder transition as shown in Fig.~\ref{fig5}(a). Whereas for the Random-sequential update, it goes continuously  from small $\langle |m| \rangle \simeq 0$ values to finite $\langle |m| \rangle$ as we increase $\beta$ Fig.~\ref{fig5} (c).
 In the Random-sequential case, $G$ goes from 2/3 (ordered state) to zero (disordered state) smoothly, as shown in Fig.~\ref{fig5}(d). To further understand the behaviour of phase transition, we calculated the order-parameter distribution $P(|m|)$ \cite{solon2013revisiting,durve2016first,bhattacherjee2015topological,singh2021bond}near the transition region $(\beta = [2.45, 2.70])$ for the Parallel and Random-sequential update case as shown in Fig. ~\ref{fig6}(a-b) respectively. For the Parallel update $P(|m|)$, shows the bimodality (phase coexistence) whereas it always unimodal for the random case and peak of the $P(|m|)$ continuously shifts to lower values of $|m|$ as we decrease $\beta$. Hence, we can say that the transition from order-to-disorder is the discontinuous type for the Parallel update and continuous type for the Random-sequential update.\\

\begin{figure}[h]
	\includegraphics[width=0.7\textwidth]{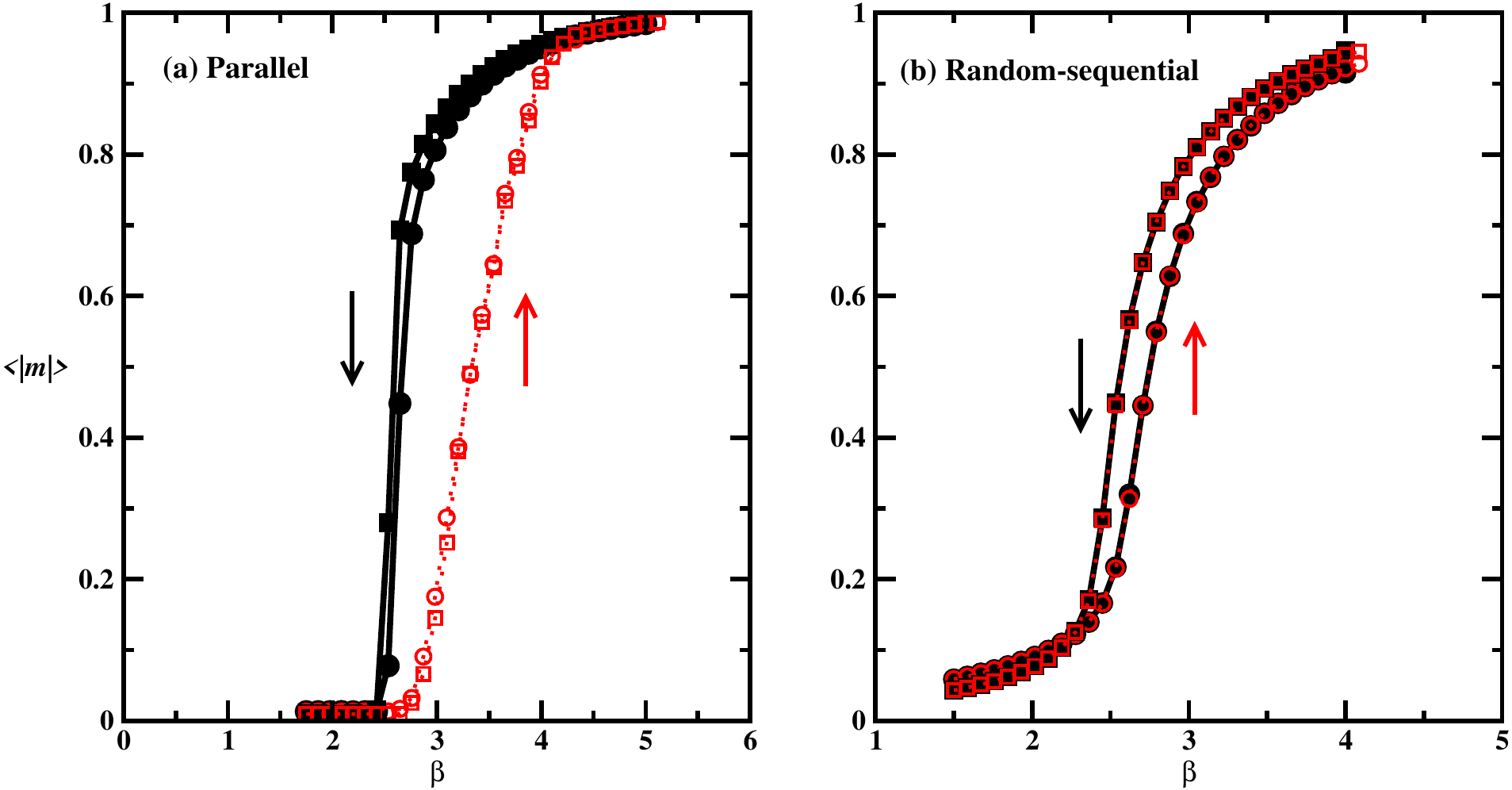}
	\caption{(color online) Hysteresis in the Parallel and Random-sequential update rule for $N = 500$ (circle) and  $1000$ (square) spins.  The order parameter $m$ vs. inverse temperate $\beta$ for ramp rate 3.6 x $10^{-6}$. The open and solid symbols are for the ramp-up and down rates, respectively. All other parameters are the same as in Fig.~\ref{fig2}.
	}
	\label{fig7}
\end{figure}

We further distinguish the difference in the nature of phase transition for the two cases by showing the presence of a hysteresis loop in the system. The presence of hysteresis is another signature of the first-order transition \cite{chate2008collective,durve2016first}. The instantaneous order parameter displays hysteresis if the control parameter is ramped up and down across the transition point at a modest (constant) ramp rate. The area of the loop varies depending on the ramp rate. In Fig.~\ref{fig7}(a-b), we show the hysteresis for $N = 500, 1000$ spins in the order parameter and the inverse temperature with a ramp rate of $3.6 $ x $10^{-6}$ per unit time for both the update rules. The open symbols are for the ramp-up and the solid filled for the ramp-down case. The two symbols, circle and square, are for two values of $N = 500$ and $1000$, respectively.  In the case of Parallel update, we get the finite area in the hysteresis loop (between ramp-up and down), but in random sequential, both curves coincide, so no area is enclosed.  This indicates that Parallel update gives us the discontinuous transition while later gives the continuous transition. The presence of finite area for the hysteresis loop is again due to the coexistence of ordered and disordered state near the critical point for the Parallel update.

\section{Conclusions}\label{IV}
We have studied flocking in one dimension using a collection of active Ising spins 
moving in the unit interval. We find that the dynamical properties of the system, both 
transient and steady state, intrinsically depend on the nature of the update rules in the system. The magnitude of the orientation fluctuations in the disordered state of the spin system is 
greater for random-sequential updates than for parallel updates. In the state of long-range order, 
the flocks alternate between the allowed orientations for the two update 
rules, the frequency of which  depends on the type of update. For the fixed set of system parameters, the system with Parallel update is less 
alternating in comparison to its Random-sequential counterpart. The density fluctuations are bimodal for the Parallel update, whereas it is unimodal for the Random sequential update. Further, we find that the differences in the density fluctuations also reflect in the nature of the transition from disorder to long-range order, with discontinuous for the
Parallel update and continuous for the Random-sequential update. 
The differences arise due to intrinsic randomness in the Random-sequential update, which makes such an evolution asynchronous as opposed to the Parallel update, which is 
inherently synchronous. The present study has implications for the current understanding 
of collective motion in one dimension and opens a similar question for other active matter systems in higher dimensions. 

\section{acknowledgement}
A.K. and S.M. thank PARAM Shivay for
the computational facility under the National Supercomputing
Mission, Government of India, at the Indian Institute of
Technology (BHU) Varanasi and also I.I.T. (BHU) Varanasi computational facility. A. K. thanks PMRF, INDIA for the research fellowship. S. M. thanks DST, SERB (INDIA), Project No.
CRG/2021/006945 and MTR/2021/000438 for partial financial support.\\

%\section{Leader}
%The Parallel update rule is synchronous, whereas Random-sequential is inherently asynchronous. Due to its inherent randomness, the system reverses its direction of motion frequently, while in Parallel update, the system occasionally reverses its direction of motion, and the reversal time is quite large. To check the response of the system to a perturbation in the dynamics of the flock for both cases, To introduce the perturbation(leader)  when the system is in the steady state, we randomly chose a spin, fixed its orientation, and let the system evolve for a long time. We have found that in the Parallel update case, where the reversal was quite less, it increases, and the system settles down in the direction of the leader. But in the Random-sequential update case, the system's behavior is similar to the scenario when there is no leader. So we can conclude that since the Random-sequential update rule has inherent randomness it is habitual to the fluctuation. However, due to the synchronous nature of the update, the parallel update rule is very sensitive to fluctuation. 

\end{document}